\documentclass[11pt]{amsart}

\usepackage{amsmath}
\usepackage{amssymb}
\usepackage{graphicx}

\theoremstyle{definition}

\numberwithin{equation}{section}

\title[Adventures in Two Time-Dependent Billiards]{Exploring the Fascinating World of Bouncing Balls: Adventures in Two Time-Dependent Billiards}

\author[Edson Denis Leonel]{Edson Denis Leonel}

\address{Departamento de F\'isica, Universidade Estadual Paulista (UNESP), Av. 24A, 1515, 13506-900 SP, Brazil}
\email{{\tt edson-denis.leonel@unesp.br}}


\subjclass[2010]{????????}

\begin{document}

\begin{abstract}
This paper explores two instances where dissipation plays a crucial role in curbing the unbounded energy growth of particles in time-dependent billiards. The first example involves an elliptical-like billiard with inelastic collisions between the particle and the boundary. This scenario introduces a fractional loss of energy upon collision, resulting in the suppression of unbounded energy growth. The second example examines an oval-like billiard, where a particle undergoes continuous energy reduction due to a viscous drag force. Both findings emerged in 2009 from collaborative research with Lyonia Bunimovich during a visit to the Georgia Institute of Technology. 
\end{abstract}

\maketitle


\section{Introduction}

In 2009, my initial encounter with the esteemed mathematician Leonid Bunimovich, affectionately known as Lyonia, occurred during a billiard conference held in \'Aguas de Lindoia, Brazil. This remarkable event marked the commencement of a transformative journey that significantly influenced both my scientific career and personal life. The fate of this meeting was facilitated through a collaboration with Alexander (Sasha) Loskutov, who 2008 received support from FAPESP (the scientific research agency of the State of S\~ao Paulo, Brazil) to visit my research group in Rio Claro. This connection laid the foundation for my introduction to Lyonia Bunimovich and paved the way for our subsequent collaborative endeavors.

At the 2009 conference, an opportunity arose to apply for a visit to Lyonia Bunimovich's group in Atlanta, facilitated by the esteemed Fulbright agency. This successful application led to a three-month collaboration with a true luminary in the field of dynamical systems and billiards. Throughout this productive collaboration, we authored two pivotal papers that impacted the scientific community, which I discuss shortly them here.

One of these papers explored the suppression of Fermi acceleration in time-dependent elliptical billiards due to inelastic collisions, published in Physical Review Letters \cite{re1}. In essence, we demonstrated that inelastic collisions resulted in a fractional energy loss upon each collision, challenging the notion of unbounded energy growth and, by extension, questioning the robustness of Fermi acceleration.

In a second paper published in Physical Review E \cite{re2}, we investigated the dynamics of particles moving within a time-dependent oval-like billiard, subject to a viscous drag force acting along the particle's trajectory. Once again, we concluded that Fermi acceleration is not a robust phenomenon. These publications have been widely read and cited within the scientific community, underscoring their significance in our field.

Our collaboration extended beyond research papers and conferences. During my visit to Atlanta, Lyonia welcomed me with open arms, leading to fruitful discussions and intellectually stimulating conversations. I was honored with the opportunity to deliver a comprehensive seminar at the mathematics school, sharing insights and knowledge with Lyonia's colleagues and students. In a more personal setting, I had the privilege of sharing a dinner with his family, further fortifying the bonds of friendship and collaboration.

Reflecting on the invaluable experiences and knowledge gained through my association with Lyonia Bunimovich, I am profoundly grateful for the opportunity to work alongside such an esteemed scientist and mentor. I extend my warmest wishes for his 75th birthday and many more years of good health, happiness, and continued brilliance in the field of dynamical systems. Lyonia Bunimovich has made an indelible mark on the scientific community and remains a source of inspiration for us all.

\section{A time-dependent elliptical-like billiard}

In this section, I will discuss some of the results found for the suppression of Fermi acceleration observed in a time-dependent elliptical-like billiard. The section was entirely based on our paper published in Physical Review Letters \cite{re1}.

We start by mentioning that Fermi acceleration (FA) is a phenomenon where a classical particle acquires unlimited energy upon collisions with a heavy and moving wall. This was the original idea considered by Enrico Fermi \cite{re3}, who assumed that the enormous energy of cosmic particles could come from interactions with moving magnetic clouds.

The phenomenon can be measured in simulations in billiards, particularly those with a moving boundary. Applications of billiards to physical problems include superconducting \cite{re4} and confinement of electrons in semiconductors by electric potentials \cite{re5,re6}, ultra-cold atoms trapped in a laser potential \cite{re7,re8,re9,re10}, mesoscopic quantum dots \cite{re11}, reflection of light from mirrors \cite{re12}, waveguides \cite{re13,re14}, and microwave billiards \cite{re15,re16}, among many others.

If the boundary is time-dependent, the Loskutov-Ryabov-Akinshin (LRA) conjecture \cite{re17} assumes that chaotic dynamics for a billiard with a static boundary is a sufficient condition to produce FA if a time perturbation of the boundary is introduced. This conjecture was confirmed in many models \cite{re18,re19}. A group led by Peter Schmelcher in Germany published a result \cite{re21} with a specific perturbation in the boundary of an integrable elliptical billiard producing tunable FA. The result discussed in \cite{re21} was a break of two critical concepts: (i) it was expected \cite{re22} that an elliptical billiard, which is integrable for a static boundary and therefore demonstrates the most regular dynamics, does not exhibit FA, and; (ii) since the static version of the elliptical billiard does not have chaotic dynamics, it then concluded that the LRA conjecture \cite{re17} must be extended.

My contribution, together with Lyonia, in this topic was to show that the mechanism producing FA in a time-dependent elliptical-like billiard (see Ref. \cite{re1}) is broken by non-elastic collisions (fractional loss of energy upon collision). Moreover, the destruction is observed for minimal dissipation. It then proves FA is not a robust phenomenon. To do that, we investigated the dynamics of an ensemble of non-interacting particles in a time-dependent elliptical-like billiard. We showed a set of initial conditions chosen along the separatrix curve of the billiard with a static boundary leads the particles to exhibit unbounded energy growth. We noticed that the LRA conjecture can be extended to the existence of a heteroclinic orbit in the phase space instead of a set with chaotic dynamics. Ref. \cite{re21} argues that the mechanism responsible for producing FA is the successive crossings by the particle of a neighborhood of a separatrix curve in the static case. It, of course, turns into a stochastic layer under time perturbation of the boundary. Such crossings change the dynamics of the particle from rotation to libration (or vice-versa) and cause the kinetic energy of the particle to fluctuate. These fluctuations increase with time, leading to anomalous diffusion and FA. Lyonia and I demonstrated that inelastic collisions of the particle with the boundary break down the mechanism of FA and, therefore, suppress the unlimited energy gain of the bouncing particle. The dissipation stops the successive crossings of the particle of the stochastic layer, thus acting on the core of the mechanism responsible for producing the diffusion in energy. This suppression confirms a conjecture \cite{re23} for the suppression of FA in 2-D billiards under inelastic collisions.

The model under consideration involves a collection of classical, non-interacting particles confined within a closed domain with a time-varying boundary. The polar coordinates equation for the boundary is expressed as:
\begin{equation}
R(\theta, e, a, t) = \frac{1 - e^2[1 + a\cos(t)]^2}{1 + e[1 + a\cos(t)]\cos(q\theta)}.
\label{eq1}
\end{equation}

Here, $e$ represents the ellipse's eccentricity, $q \ge 1$ is an integer, $a$ denotes the amplitude of time perturbations, $\theta$ is the angular coordinate and $t$ is time. The static case, where $a=0$ and $q=1$, corresponds to the classical scenario.

Two conserved quantities exist in the system: the kinetic energy of the particle \cite{re24} and the angular momentum about the two foci \cite{re25}. The latter is described by:
\begin{equation}
F(\alpha, \theta) = \frac{\cos^2(\alpha) - e^2\cos^2(\phi)}{1 - e^2\cos^2(\phi)},
\label{eq1a}
\end{equation}
where $\phi = \arctan(Y^{\prime}(\theta_n,t_n)/X^{\prime}(\theta_n,t_n))$, $X^{\prime} = dX/d\theta$, and $Y^{\prime} = dY/d\theta$.

\begin{figure}[t]
\centerline{\includegraphics[width=0.6\linewidth]{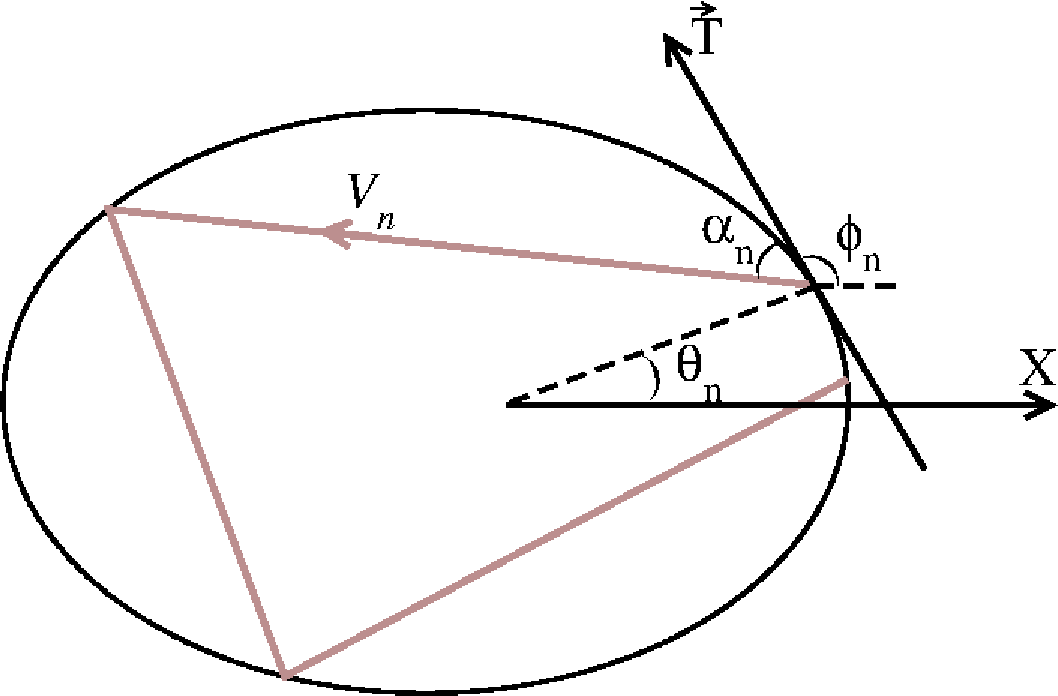}}
\caption{A typical illustration of an elliptical billiard.}
\label{Fig1}
\end{figure}

An implicit 4-D mapping governs the dynamics of the model. Traditionally, investigations start with the initial condition $(\theta_n, \alpha_n, V_n, t_n)$, where $\alpha_n$ represents the angle between the particle's trajectory and the tangent line to the boundary at $\theta_n$, $V_n>0$ is the particle's velocity, and $t_n$ is the moment of the $n^{th}$ collision with the boundary. Figure \ref{Fig1} illustrates the corresponding coordinate angles for a typical orbit in the elliptical billiard.

From the initial condition, the particle's dynamics are described by $X(t) = X(\theta_n,t_n) + |\vec{V}_n|\cos(\alpha_n+\phi_n)(t-t_n)$ and $Y(t) = Y(\theta_n,t_n) + |\vec{V}n|\sin(\alpha_n+\phi_n)(t-t_n)$. The angular coordinate $\theta_{n+1}$ is obtained by following the trajectory of the particle until the time $t = t_n + \Delta t$, where $\Delta t$ satisfies the equation:
\begin{equation}
\sqrt{X^2(\Delta t) + Y^2(\Delta t)} = \frac{1 - e^2[1 + a\cos(\Delta t)]^2}{1 + e[1 + a\cos(\Delta t)]\cos(q\theta)}.
\label{eq2}
\end{equation}

The time for a further impact is given by:
\begin{equation}
t_{n+1} = t_n + \sqrt{\frac{[X(\theta_{n+1}) - X(\theta_n)]^2 + [Y(\theta_{n+1}) - Y(\theta_n)]^2}{|\vec{V}_n|}}.
\label{eq3}
\end{equation}

Upon collision with the boundary, the reflection law is applied:
\begin{eqnarray}
\vec{V}^{\prime}_{n+1}\cdot \vec{T}_{n+1} &=& \vec{V}^{\prime}_n\cdot \vec{T}_{n+1}, \label{eq4}\\
\vec{V}^{\prime}_{n+1}\cdot \vec{N}_{n+1} &=& -\gamma\vec{V}^{\prime}_n\cdot \vec{N}_{n+1},
\label{eq5}
\end{eqnarray}
where $\vec{T}$ and $\vec{N}$ are the unit tangent and normal vectors, and $\gamma \in [0,1]$ is the restitution coefficient. The case of $\gamma=1$ corresponds to elastic collisions.

The components of the particle's velocity after collision are given by:
\begin{eqnarray}
\vec{V}_{n+1}\cdot \vec{T}_{n+1} &=& |\vec{V}_n|[\cos(\alpha_n+\phi_n)\cos(\phi_{n+1})] \nonumber \\
&+& |\vec{V}_n|[\sin(\alpha_n+\phi_n)\sin(\phi_{n+1})], \label{eq6} \\
\vec{V}_{n+1}\cdot \vec{N}_{n+1} &=& -\gamma|\vec{V}_n|[\sin(\alpha_n+\phi_n)\cos(\phi_{n+1})] \nonumber \\
&-& \gamma|\vec{V}_n|[-\cos(\alpha_n+\phi_n)\sin(\phi_{n+1})] \nonumber \\
&+& (1+\gamma)\frac{dR(t)}{dt}[\sin(\theta_{n+1})\cos(\phi_{n+1})] \nonumber \\
&-& (1+\gamma)\frac{dR(t)}{dt}[\cos(\theta_{n+1})\sin(\phi_{n+1})],
\label{eq7}
\end{eqnarray}
where $\frac{dR}{dt}$ is the velocity of the moving boundary at the $(n+1)$-th collision. The particle's velocity after collision is given by:
\begin{equation}
V_{n+1} = \sqrt{(\vec{V}_{n+1}\cdot\vec{T}_{n+1})^2 + (\vec{V}_{n+1}\cdot\vec{N}_{n+1})^2}.
\label{eq8}
\end{equation}

The coordinate angle $\alpha_{n+1}$ is determined by:
\begin{equation}
\alpha_{n+1} = \arctan\left(\frac{\vec{V}_{n+1}\cdot\vec{N}_{n+1}}{\vec{V}_{n+1}\cdot\vec{T}_{n+1}}\right).
\label{eq9}
\end{equation}

\begin{figure}[t]
\centerline{\includegraphics[width=0.8\linewidth]{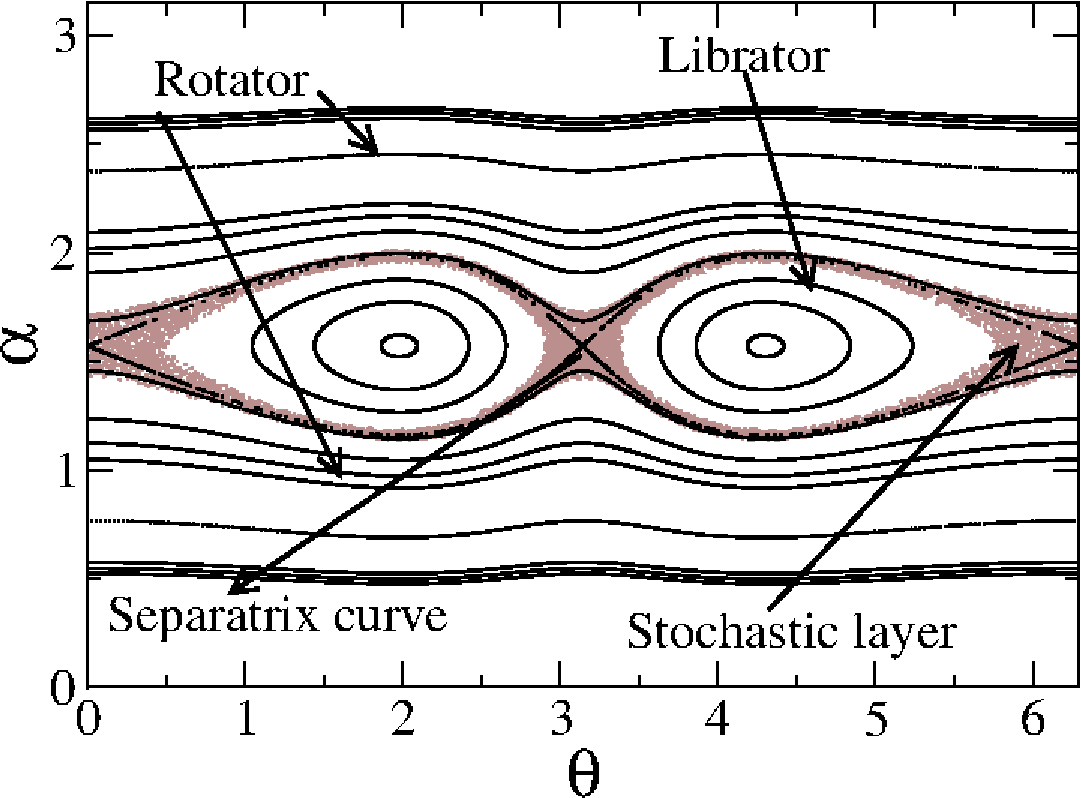}}
\caption{Phase space plot for the static billiard overlapped with a stochastic layer created by time perturbations. Control parameters: static case $e=0.4$, $q=1$; time-dependent perturbation $e=0.4$, $a=0.01$, $\gamma=1$ with $V_0=1$ and $10^4$ collisions with the edge.}
\label{Fig2}
\end{figure}

Figure \ref{Fig2} displays a phase space plot for the static billiard overlaid with a stochastic layer generated by time perturbations. The control parameters considered were $e=0.4$, $q=1$ for the static case, and $e=0.4$, $a=0.01$, $\gamma=1$ with $V_0=1$ for time-dependent perturbation with $10^4$ collisions with the edge.

\begin{figure}[t]
\centerline{\includegraphics[width=0.7\linewidth]{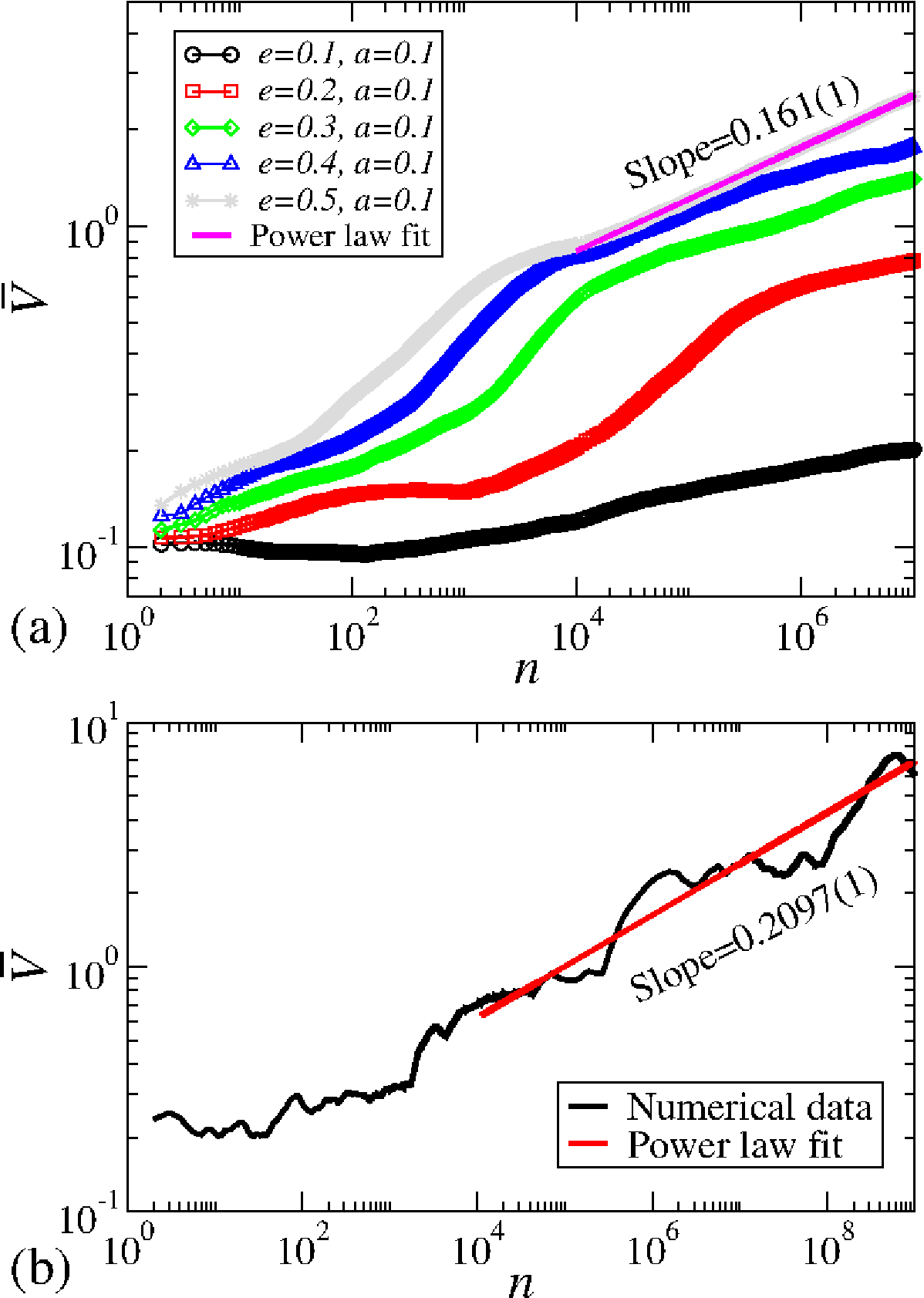}}
\caption{(a) Plot of ${\bar V}\times n$ for an ensemble of $100$ particles for $q=1$. Control parameters are labeled in the figure. (b) Plot of ${\bar V}\times n$ for a single particle. Control parameters: $e=0.5$, $a=0.1$, $\gamma=1$. The slope obtained is $0.2098(1)$.}
\label{Fig3}
\end{figure}

Figure \ref{Fig3} presents the particle's average velocity as a function of $n$ for different control parameters. Initial conditions were $\alpha_0=\pi/2$, $\theta_0=\pi$, corresponding to the location of the heteroclinic point along the separatrix curve for the static boundary with $q=1$, $V_0=10^{-1}$, and $100$ uniformly distributed $t_0\in[0,2\pi]$. The control parameters are labeled in the figure.

After an initial transient and a regime of fast growth marked by turbulence in the behavior of $F$ (see Fig. \ref{Fig5}(b)), the curves stabilize into a constant growth regime. A power-law fitting for the control parameters $e=0.5$ and $a=0.1$ gives a slope of $0.161(1)$, indicating energy diffusion. Simulations considered $10^7$ collisions with the boundary. Figure \ref{Fig3}(b) shows a simulation for a single initial condition considering $10^9$ collisions with the boundary. A power-law fitting gives $0.2097(1)$, confirming the presence of FA. These results extend the LRA conjecture \cite{re17} and replace chaotic dynamics for the static boundary with the existence of a heteroclinic orbit in the phase space. The introduction of time perturbations turns the separatrix curve into a stochastic layer \cite{re21}, leading to FA.

\begin{figure}[t]
\centerline{\includegraphics[width=0.7\linewidth]{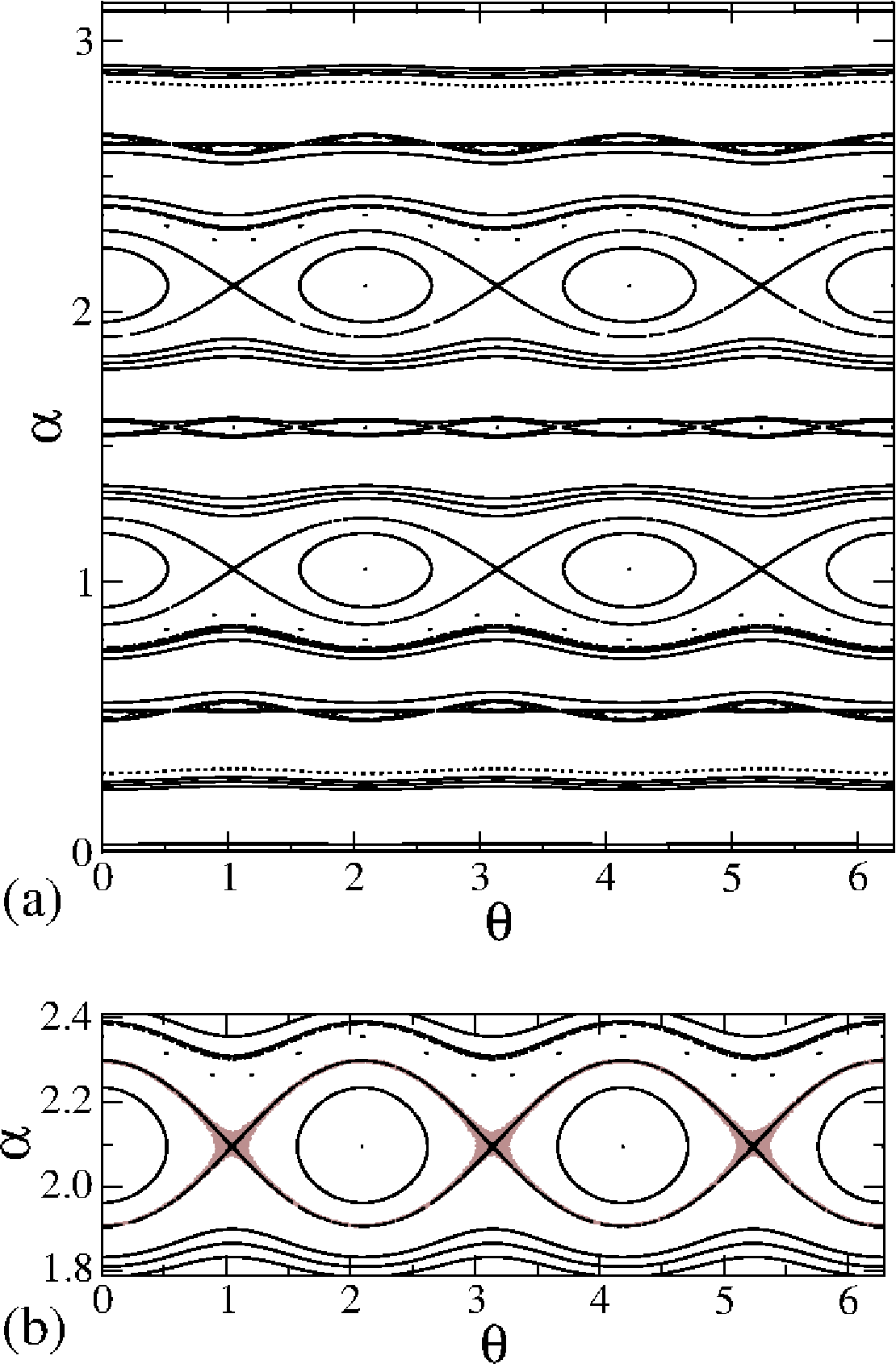}}
\caption{(a) Phase space plot for $q=3$ and $e=0.01$. (b) Zoom-in of the upper period, three chains overlapped by a stochastic layer created by the time dependence. Simulations used $a=0.01$.}
\label{Fig4}
\end{figure}

\begin{figure}[t]
\centerline{\includegraphics[width=0.7\linewidth]{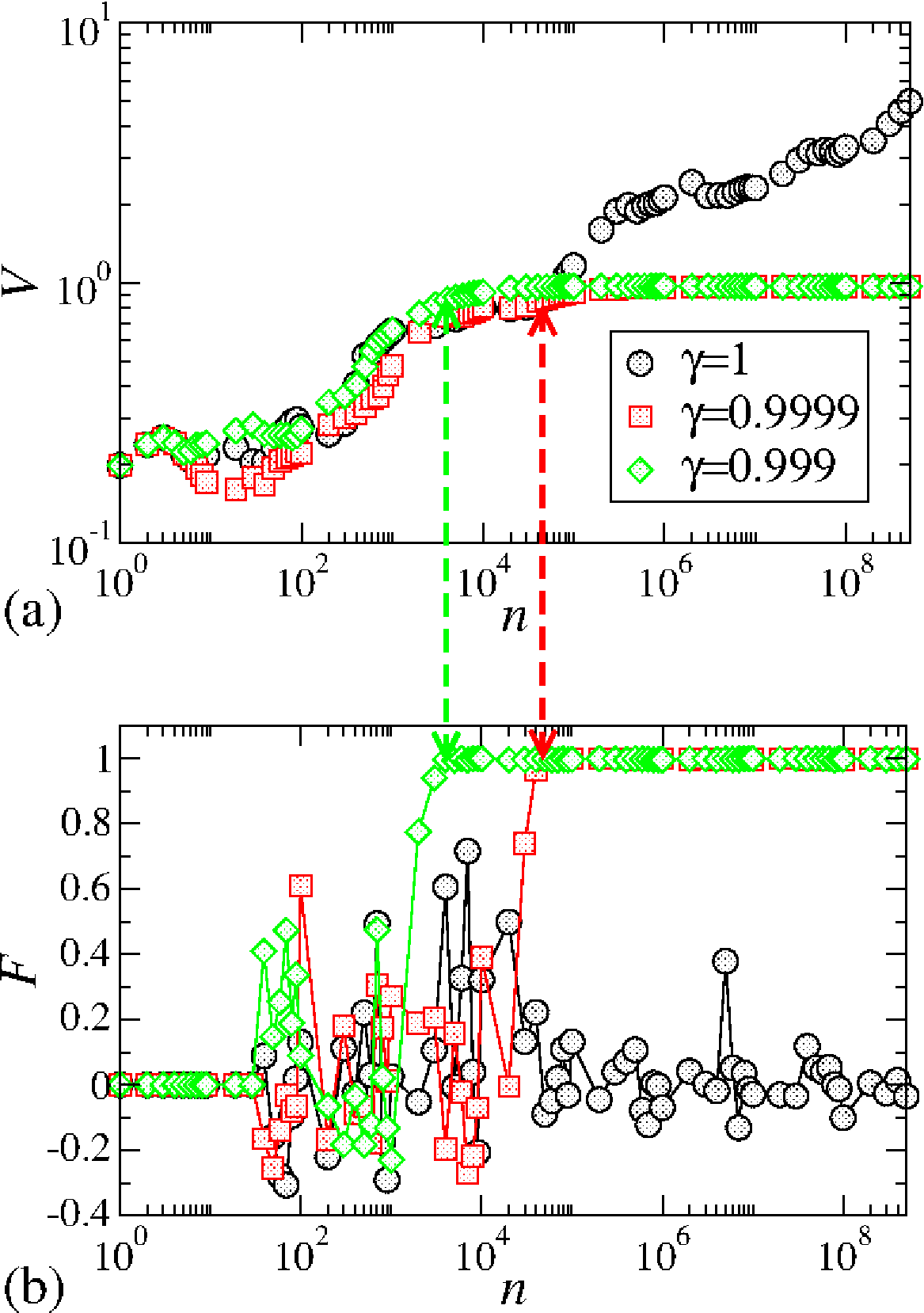}}
\caption{(a) Plot of $V_n\times n$ for different restitution coefficients, as marked in the figure. (b) Plot of $F_n\times n$ for
the same control parameters as in (a).}
\label{Fig5}
\end{figure}

We demonstrate that the observed behavior for the elliptical domain with $q=1$ extends to other boundary shapes. If very thin stochastic layers exist, the introduction of time dependence enlarges them, leading the particle to exhibit FA. As an illustration, consider $q=3$. For the static case with $a=0$, the control parameter $e_c=1/(q^2-1)$ marks a change when the boundary exhibits non-concave pieces. Invariant spanning curves in the phase space are destroyed for any $e\ge e_c$. Figure \ref{Fig4}(a) shows the phase space for $q=3$, featuring two symmetric chains of period three orbits separated by a thin stochastic layer. Overlapping with the stochastic layers for a time-dependent boundary in Fig. \ref{Fig4}(b) suggests the occurrence of FA.

To conclude this section, let us discuss the effect of inelastic collisions when $\gamma<1$. Upon collision, there is a fractional energy loss, significantly altering the particle's dynamics. The particle traverses the stochastic layer for a while, then escapes, becoming trapped in either librator or rotator orbits. Figure \ref{Fig5}(a) shows the velocity of the particle as a function of $n$ for $q=1$ and three different damping coefficients: $\gamma=1$ (non-dissipative case); $\gamma=0.9999$; and $\gamma=0.999$. 
Even for a small dissipation, the regime of energy growth is interrupted, leading the particle's energy to reach a constant plateau. Figure \ref{Fig5}(b) demonstrates, in a log-linear plot, the time evolution of the observable $F$ (see Eq. (\ref{eq1a})). For $\gamma=1$, the value of $F$ fluctuates around $0$ up to $5\times 10^8$ collisions. The dynamics for two values of $\gamma<1$ show trapping in rotator orbits after a few hundred collisions. However, the particle can also evolve towards librator orbits. These results confirm that FA is suppressed due to the breakdown of the mechanism producing FA. Since the mapping changes minimally with inelastic collisions, this evidence suggests that FA is not a robust phenomenon.

These results demonstrate that FA is not a robust phenomenon and is suppressed when the particle experiences fractional energy loss upon collisions with the boundary.

\section{A time-dependent oval-like billiard}

In this section we discuss our second finding, as published in Physical Review E \cite{re2}. We consider a dissipative oval-like billiard with a boundary moving periodically in time. The dissipation acts along of the particle's trajectory and is assumed to be proportional to a power of the velocity $V$ of the particle. The three specific types of power laws we discussed were: (i) $F\propto -V$; (ii) $F\propto -V^2$ and (iii) $F\propto -V^{\delta}$ with $1<\delta<2$. If a large initial velocity is considered, case (i) shows a decay of the particle's velocity is a linear function of the number of collisions with the boundary. For case (ii), an exponential decay is observed, and for $1<\delta<2$, a power law decay is observed. We present a set of scaling hypotheses to characterize a phase transition from limited to unlimited energy gain for cases (ii) and (iii). The critical exponents obtained for the phase transition in the case (ii) are the same as those obtained for the dissipative bouncer model \cite{re26}. This proves near the transition, the two different models belong to the same class of universality. For all types of dissipation the suppression of the unlimited energy growth is observed.

The model considered consists of a classical particle confined to a domain with the boundary moving in time according to the following equation in polar coordinates
\begin{equation}
R(\theta,\epsilon,a,t,p)=1+\epsilon[1+a\cos(t)]\cos(p\theta).
\label{aeq1}
\end{equation}
Here $\epsilon$ is the amplitude of the circle's perturbation, $a$ is the amplitude of the time perturbation, $\theta$ is the angular coordinate, $t$ is the time and $p>0$ is an integer. The variation of the control parameters allow us to obtain different kinds of geometry. If $\epsilon=0$, the circular billiard is recovered. The phase space shows a foliate shape filled with invariant curves and absence of chaos \cite{re24}. If $\epsilon\ne 0$ and $a=0$, for $\epsilon<\epsilon_c=1/(p^2+1)$, the phase space contains periodic islands, invariant spanning curves corresponding to rotating orbits (invariant tori) and chaotic regions \cite{re25} while for $\epsilon\ge\epsilon_c$ all the invariant tori are destroyed \cite{re27}, resting only KAM islands. For $a\ne 0$ the particle may gain or lose energy upon collisions with the boundary. Since it is known \cite{re18,re19} that FA is
observed in driven oval-like billiards, our goal in is to investigate the effects of a drag-type force in the system.

As in the previous section, the dynamics of the model is described by an implicit 4-D mapping $T(\theta_n,\alpha_n,V_n,t_n)=(\theta_{n+1},\alpha_{n+1},V_{n+1},t_{n+1})$. We consider three different laws for the damping force acting on the particle: (i) $F=-\eta^{\prime}V$; (ii) $F=-\eta^{\prime}V^2$ and (iii) $F=-\eta^{\prime}V^{\delta}$ with $\delta\ne 1$  and $\delta\ne 2$ where $\eta^{\prime}$ is the viscosity coefficient. We assume it as a constant along the particle's trajectory. It is interesting to mention that other different kinds of forces proportional to the particle's velocity have also been considered in the literature. In particular, if the particle is moving under the action of a magnetic field \cite{re30,re31,re32}, it will move along of Larmor arcs of radius $R$ which are proportional to the particle's velocity. A property of such kind of perturbation is that time-reversal symmetry is broken \cite{re30}. 

The expressions for the mapping will be obtained for the case (i). To obtain the expression of the velocity of the particle along its trajectory, the second Newton's law of motion must be solved. After integration of $-\eta^{\prime}V=mdV/dt$ with the initial velocity $V_n>0$, we obtain $V(t)=V_n\exp[-\eta(t-t_n)]$, with $\eta=\eta^{\prime}/m$ and $t\ge t_n$. A displacement of the particle along a straight line is obtained from integration of $dr/dt=V(t)$, yielding $r(t)=V_n[1-\exp(-\eta(t-t_n))]/\eta$ for $t\ge t_n$. The coordinates of the particle are
$X(t)=R(\theta_n,t_n)\cos(\theta_n)+r(t)\cos(\phi_n+\alpha_n)$ and
$Y(t)=R(\theta_n,t_n)\sin(\theta_n)+r(t)\sin(\phi_n+\alpha_n)$ with
$\phi_n=\arctan[Y^{\prime}(\theta_n,t_n)/X^{\prime}(\theta_n,t_n)]$ and
$X^{\prime}=dX/d\theta$ and $Y^{\prime}=dY/d\theta$. Two cases may occur after the particle suffers a collision with the boundary and leaves the collision zone (region on the plane circumscribed by $R(\theta)\le R_c(\theta)=1-\epsilon (1+a)$): (a) the particle has enough energy to enter the collision zone again ($R\ge R_c$) and has another collision with the boundary or; (b) the particle does not have enough energy to reach a point of the next collision and, thanks to the dissipation, the particle stops, yielding a maximum displacement $r_{\rm max}=V_n/\eta$. The new angular coordinate $\theta_{n+1}$ is obtained, after evolving the dynamics of the particle in time, as a root of the following equation
\begin{equation}
\sqrt{X^2(t-t_n)+Y^2(t-t_n)}=1+\epsilon[1+a\cos(t-t_n)]\cos(p\theta)~,
\label{aeq2}
\end{equation}
with $t\ge t_n$. The time at a further collision is
\begin{equation}
t_{n+1}=t_n+\Delta t_n~,
\label{aeq3}
\end{equation}
where $\Delta t_n=-\ln[1-\eta r(t_c)/V_n]/\eta$, with $t_c$
representing the instant of the collision.

The reflection rules are the following
\begin{eqnarray}
\vec{V}^{\prime}_{n+1}\cdot
\vec{T}_{n+1}=\vec{V_p}^{\prime}(t_{n+1})\cdot
\vec{T}_{n+1}~,\label{aeq4}\\
\vec{V}^{\prime}_{n+1}\cdot
\vec{N}_{n+1}=-\vec{V_p}^{\prime}(t_{n+1})\cdot
\vec{N}_{n+1}~,
\label{aeq5}
\end{eqnarray}
where the upper prime denotes the velocities are measured with
respect to the moving boundary reference frame. The vectors $\vec{T}$
and $\vec{N}$ are the unit tangent and normal vectors respectively and
$\vec{V_p}(t_{n+1})$ is the velocity of the particle immediately before
the collision given by $\vec{V_p}(t_{n+1})=V_n\exp[-\eta(t_{n+1}-t_n)]$.

Based on Eqs. (\ref{aeq4}) and (\ref{aeq5}), the components of the
velocity of the particle after a collision are
\begin{eqnarray}
\vec{V}_{n+1}\cdot
\vec{T}_{n+1}&=&|\vec{V_p}(t_{n+1})|[\cos(\alpha_n+\phi_n)\cos(\phi_{n+1
})] \nonumber \\
&+&|\vec{V_p}(t_{n+1})|[\sin(\alpha_n+\phi_n)\sin(\phi_{n+1})]~,
\label{aeq6}\\
\vec{V}_{n+1}\cdot\vec{N}_{n+1}&=&-|\vec{V_p}(t_{n+1})|[
\sin(\alpha_n+\phi_n)\cos(\phi_{n+1})]\nonumber \\
&-&|\vec{V_p}(t_{n+1})|[-\cos(\alpha_n+\phi_n)\sin(\phi_{n+1})]
\nonumber\\
&+&2{{dR(t)}\over{dt}}[\sin(\theta_{n+1})\cos(\phi_{n+1})]
\nonumber \\
&-&2{{dR(t)}\over{dt}}[\cos(\theta_{n+1})\sin(\phi_{n+1})]~,
\label{aeq7}
\end{eqnarray}
where $dR/dt$ is the velocity of the moving boundary at the instant of
collision. The velocity of the particle after the $(n+1)^{th}$
collision is
\begin{equation}
V_{n+1}=\sqrt{(\vec{V}_{n+1}\cdot\vec{T}_{n+1})^2+(\vec{
V}_{n+1}\cdot\vec{N}_{n+1})^2}~.
\label{aeq8}
\end{equation}
The angle $\alpha_{n+1}$ is obtained as
\begin{equation}
\alpha_{n+1}=\arctan\left[{{\vec{V}_{n+1}\cdot\vec{N}_{n+1}}\over{\vec{V
}_{n+1}\cdot\vec{T}_{n+1}}}\right]~.
\label{aeq9}
\end{equation}

We discuss the results for the three types of dissipative forces considered.

\subsection{Results for the case $F=-\eta V$}

\begin{figure}[t]
\centerline{\includegraphics[width=0.6\linewidth]{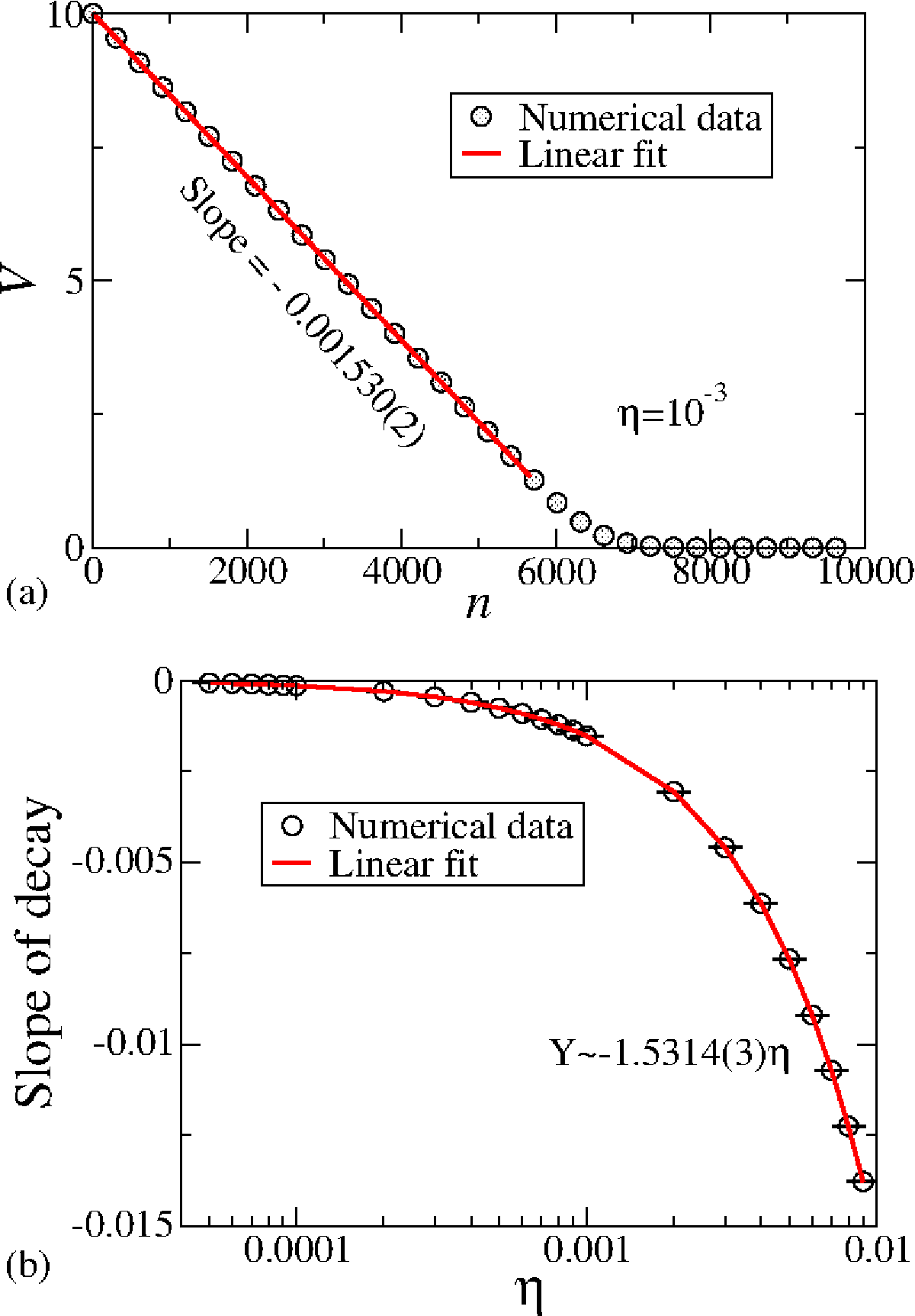}}
\caption{(a) Plot of $\bar{V}~vs.~n$. The control parameters used were $\epsilon=0.1$, $a=0.1$, $p=3$ and $\eta=10^{-3}$ and the initial velocity was $V_0=10$. (b) Linear fit of the velocity's decay slope as a function of $\eta$.}
\label{aFig2}
\end{figure}

The average velocity of the particle as a function of $n$ is shown in
Fig. \ref{aFig2}(a).
We started the simulation with the initial velocity $V_0=10$ and considering the initial angles $\alpha_0\in [0,\pi]$ and $\theta_0\in[0,2\pi]$ chosen at random in a grid of $100\times 100$ together with the initial time $t_0\in [0,2\pi]$. The control parameters were $\epsilon=0.1$, $a=0.1$, $p=3$ and $\eta=10^{-3}$. The boundary oscillates between $1-\epsilon[1+a]<R<1+\epsilon[1+a]$ and, eventually, it changes the sign of the curvature. In the static case, this change leads to the destruction of the invariant tori. The simulations are done near a critical control parameter where the invariant tori are destroyed in the static case, but some KAM islands still survive \cite{re27}. The particle's average velocity is
\begin{equation}
\bar{V}={{1}\over{M}}\sum_{j=1}^{M}{{1}\over{n}}\sum_{i=0}^{n-1}V_{j,i},
\label{aeq10}
\end{equation}
where $M=10^4$ denotes the number of different initial conditions in the ensemble.  Figure \ref{aFig2}(a) shows the average velocity decays linearly as a function of $n$. The slope of the decay obtained via a linear fit is $-0.001530(2)$ where $2\times 10^{-6}$ represents the {\it error} of the
fitting. Eventually, the particle reaches a critical velocity after
leaving the collision zone and has no energy to reach the boundary again
for the further collision. The plateau seen in Fig. \ref{aFig2}(a) for
$n>7000$ represents few orbits remaining at low velocity after the linear decay is finished before stopping completely.

We simulated the behavior of the decay of the particle's velocity for different $\eta$. The slope of the decay as a function of $\eta$ is shown in Fig. \ref{aFig2}(b). It was discussed in Ref. \cite{re28}, that the behavior of the decay of the particle's velocity for the Fermi-Ulam model (a classical particle bouncing between two rigid walls, where one of them is fixed, and the other one moves periodically in time) is linearly dependent on the control parameter $\eta$,
i.e. $V_n\propto V_0-2n\eta$. Here, the behavior of the average
velocity of the particle is $V_n\propto V_0-1.5314(3)n\eta$. The results
obtained allow us to conclude that if the particle experiences a drag
force proportional to its velocity, the phenomenon of unlimited energy
growth is suppressed.

\subsection{Results for the case $F=-\eta V^2$}
\label{sec_b}

\begin{figure}[t]
\centerline{\includegraphics[width=0.6\linewidth]{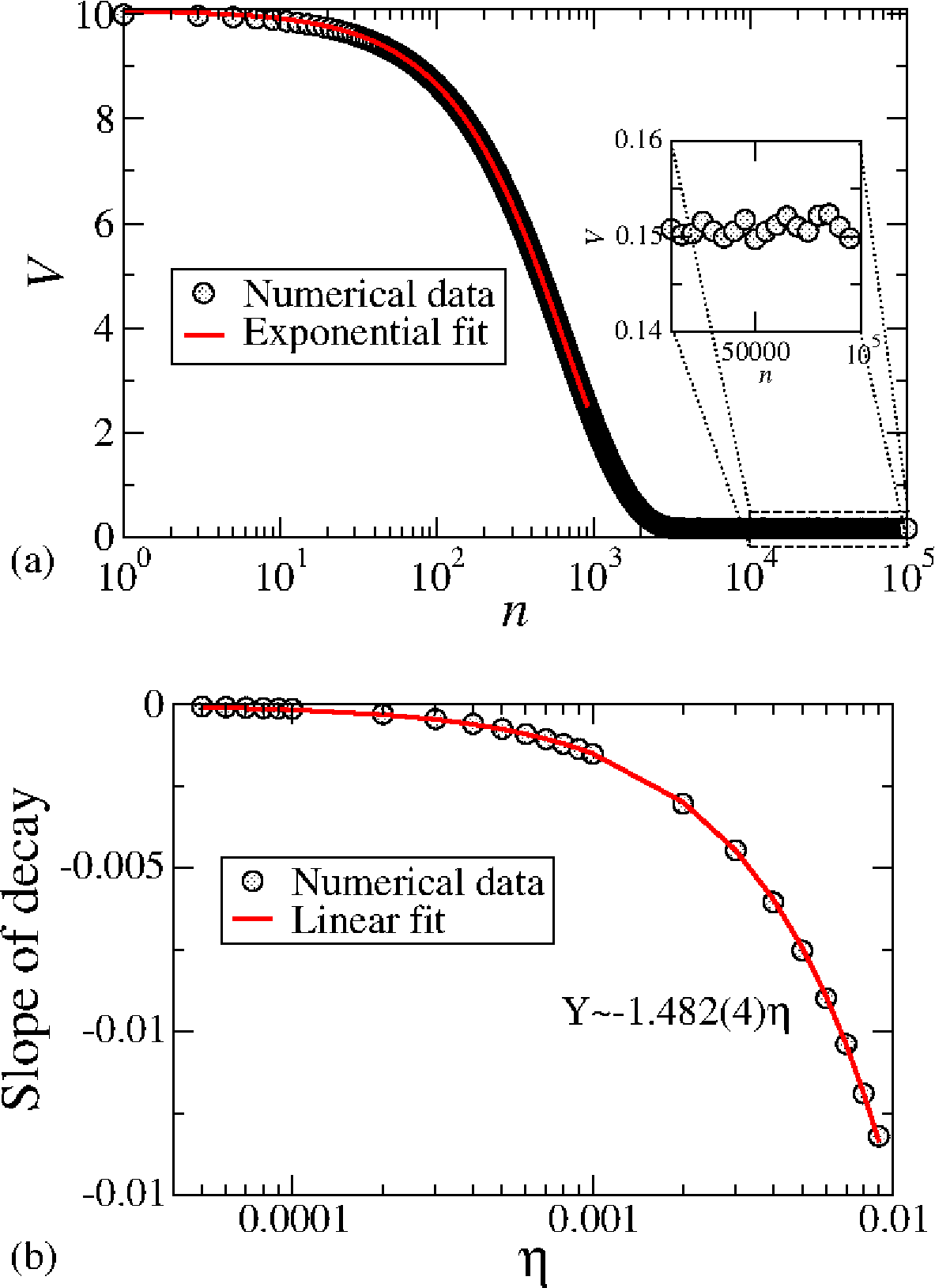}}
\caption{(a) Plot of $\bar{V}~vs.~n$ for $V_0=10$. The control parameters used were $\epsilon=0.1$, $a=0.1$, $p=3$ and $\eta=10^{-3}$. (b) Linear fit of the velocity's decay slope as a function of $\eta$.}
\label{aFig3}
\end{figure}

The equation to integrated is $-\eta V^2=dV/dt$. Considering the initial velocity as $V_n>0$, we obtain
\begin{equation}
V(t)={{V_n}\over{1+\eta(t-t_n)}}~,
\label{aeq12}
\end{equation}
with $t\ge t_n$. The integration of Eq. (\ref{aeq12}) gives
\begin{equation}
r(t)={{1}\over{\eta}}\ln[1+\eta V_n(t-t_n)]~,
\label{aeq13}
\end{equation}
for $t\ge t_n$. Updating the mapping with Eqs. (\ref{aeq12}) and (\ref{aeq13}), we follow the trajectory of the particle. The behavior of $\bar{V}$ as function of $n$ is shown in Fig. \ref{aFig3}(a).
Considering a large initial velocity $V_0=10$, the particle's velocity decays exponentially for a short time, and, after reaching a characteristic crossover $n_c$, the velocity bends towards a regime of a constant plateau. Contrary to the case of $F\propto -V$, the dissipation does not stop the particle since $r(t)$ is a function that grows in time. The particle stays moving, and the dissipation decreases as the particle's velocity diminishes. If the particle has escaped the
collision zone, it enters the zone again and suffers another collision with the boundary. An exponential fitting for the decay shown in Fig. \ref{aFig3}(a) gives that $\bar{V}=V_0\exp[-0.00153(1)n]$ for $\eta=10^{-3}$. Figure \ref{aFig3}(b) shows the behavior of the decay slope as a function of the control parameter $\eta$. A linear fitting gives the slope $\propto -1.482(4)\eta$. As discussed in
\cite{re29}, the velocity decay in the Fermi-Ulam model was
obtained analytically as $V_n=V_0\exp[-2n\eta]$. However, the time-dependent oval billiard is $V_n=V_0\exp[-1.482(4)n\eta]$. The decay obtained for the oval billiard is slower for cases (i) and (ii).

\begin{figure}[t]
\centerline{\includegraphics[width=0.6\linewidth]{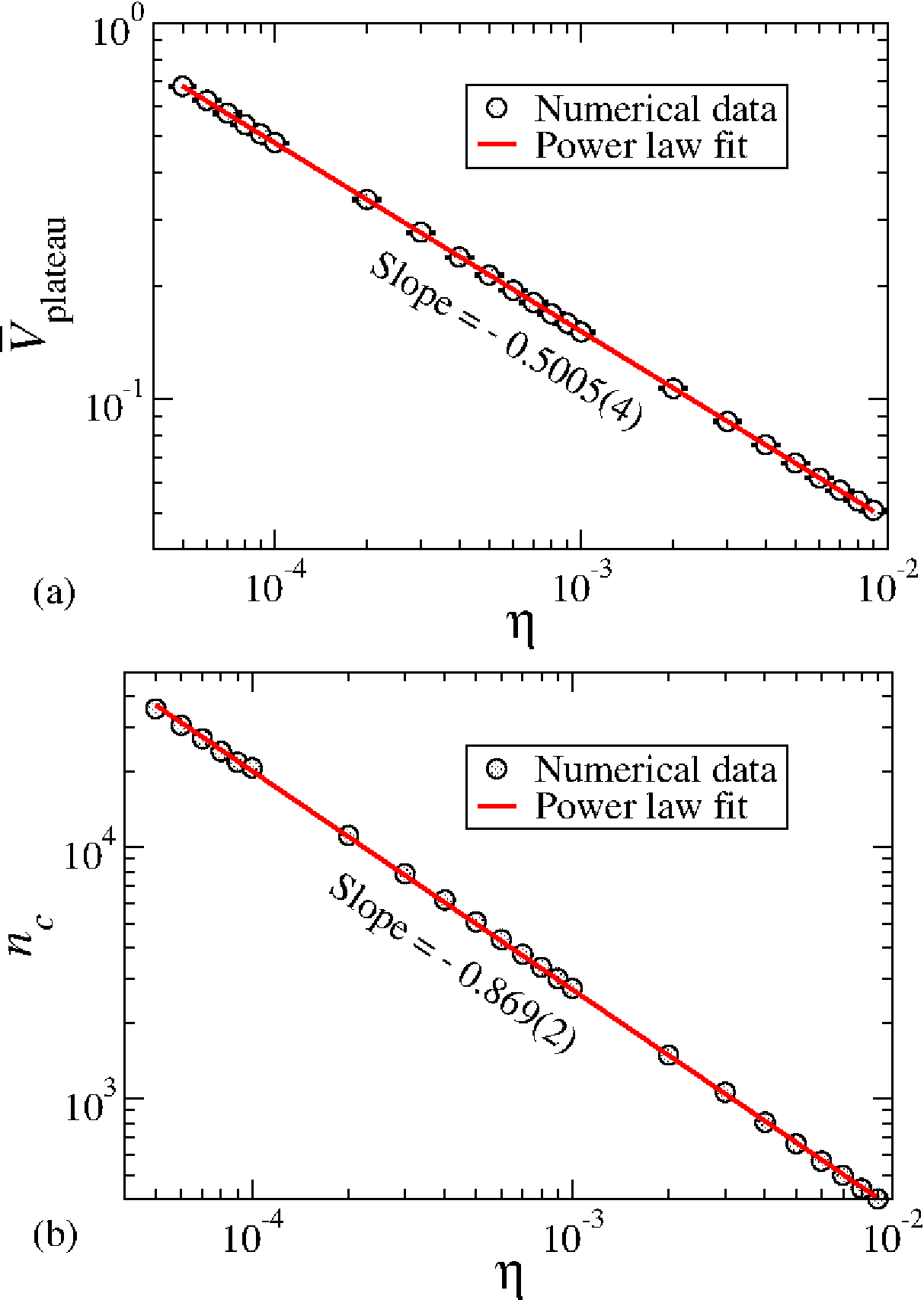}}
\caption{(a) Plot of the average velocity along the plateau for
large $n$ as a function of the control parameter $\eta$. The control
parameters were $\epsilon=0.1$, $a=0.1$ and $p=3$. (b) Plot of
$n_c$ as a function of $\eta$.}
\label{aFig4}
\end{figure}

Let us now describe the behavior of the average velocity for large values of $n$, i.e., along the constant plateau. The zoom-in in Fig. \ref{aFig3}(a) shows a few points along the constant plateau. They fluctuate around an average value, keeping larger than zero. An immediate question that arises is: what happens to the average velocity along the plateau when the strength of the dissipation is changed? If a reduction takes place, it is
expected that dissipation affects the FA less, and the constant plateau should increase. It happens! The behavior of
$V_{\rm plateau}$ as function of $\eta$ is shown in Fig. \ref{aFig4}(a).
A power law fitting gives $V_{\rm plateau}\propto \eta^{-0.5}$ leading to a divergence of the velocity as the control parameter $\eta\rightarrow 0$. This recovers FA. It is also a smooth transition from suppression to production of FA. Figure \ref{aFig4}(b) shows the behavior of the crossover iteration number $n_c$, which is the number of collisions with the boundary needed to change the regime of decay to the regime of constant velocity, for an initial velocity $V_0=10$. 

\begin{figure}[t]
\centerline{\includegraphics[width=0.6\linewidth]{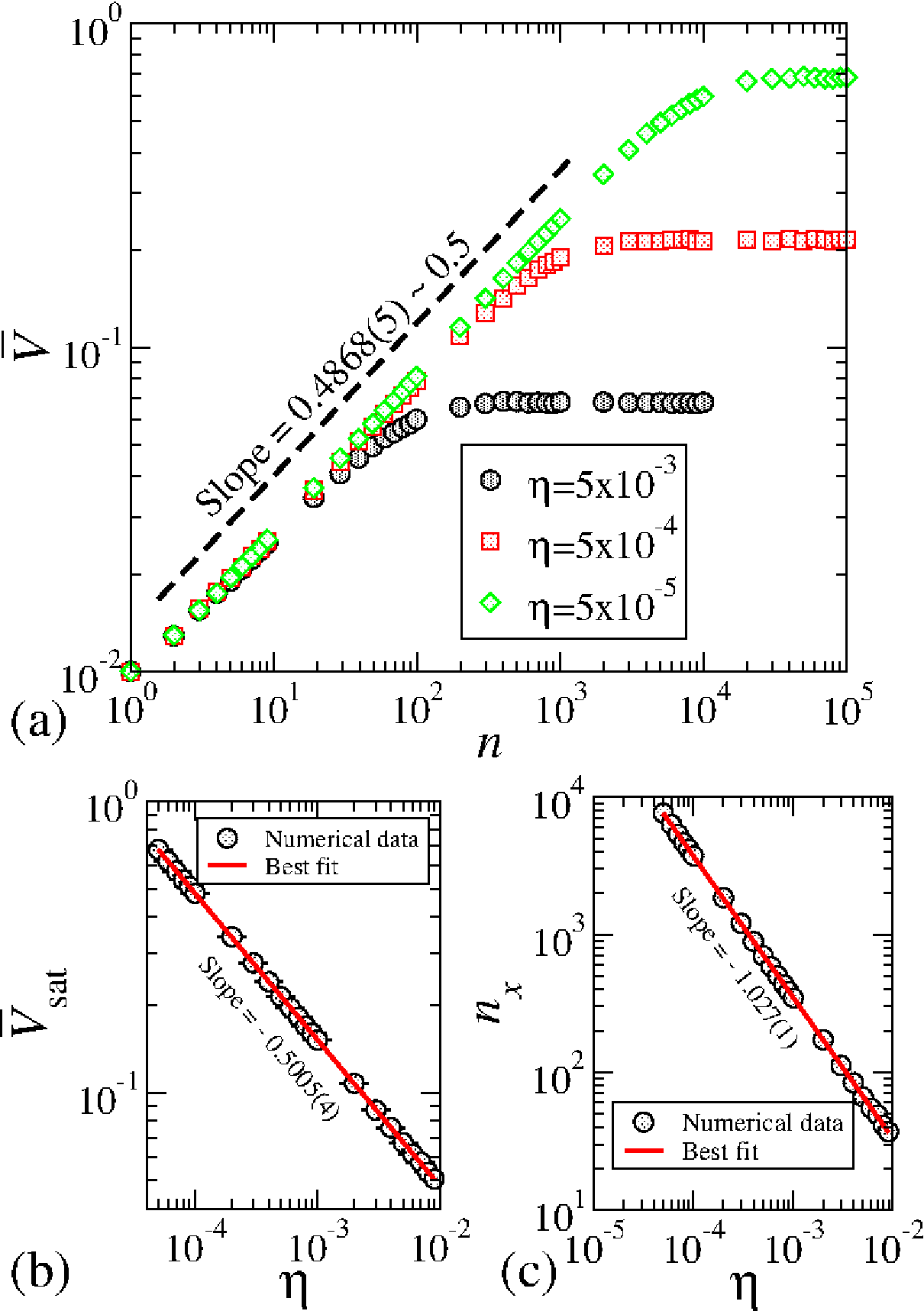}}
\caption{(a) Plot of the average velocity as function of $n$ for three different control parameters $\eta$, as shown in the figure. The initial velocity was $V_0=10^{-2}$ for the control $\epsilon=0.1$, $a=0.1$ and $p=3$. (b) Plot of $\bar{V}_{\rm sat}\times\eta$. A power law fitting yields the slope $\alpha=-0.5005(4)$. (c) Plot of $n_x\times\eta$. The slope obtained is $z=-1.027(1)$.}
\label{aFig5}
\end{figure}

\begin{figure}[t]
\centerline{\includegraphics[width=0.7\linewidth]{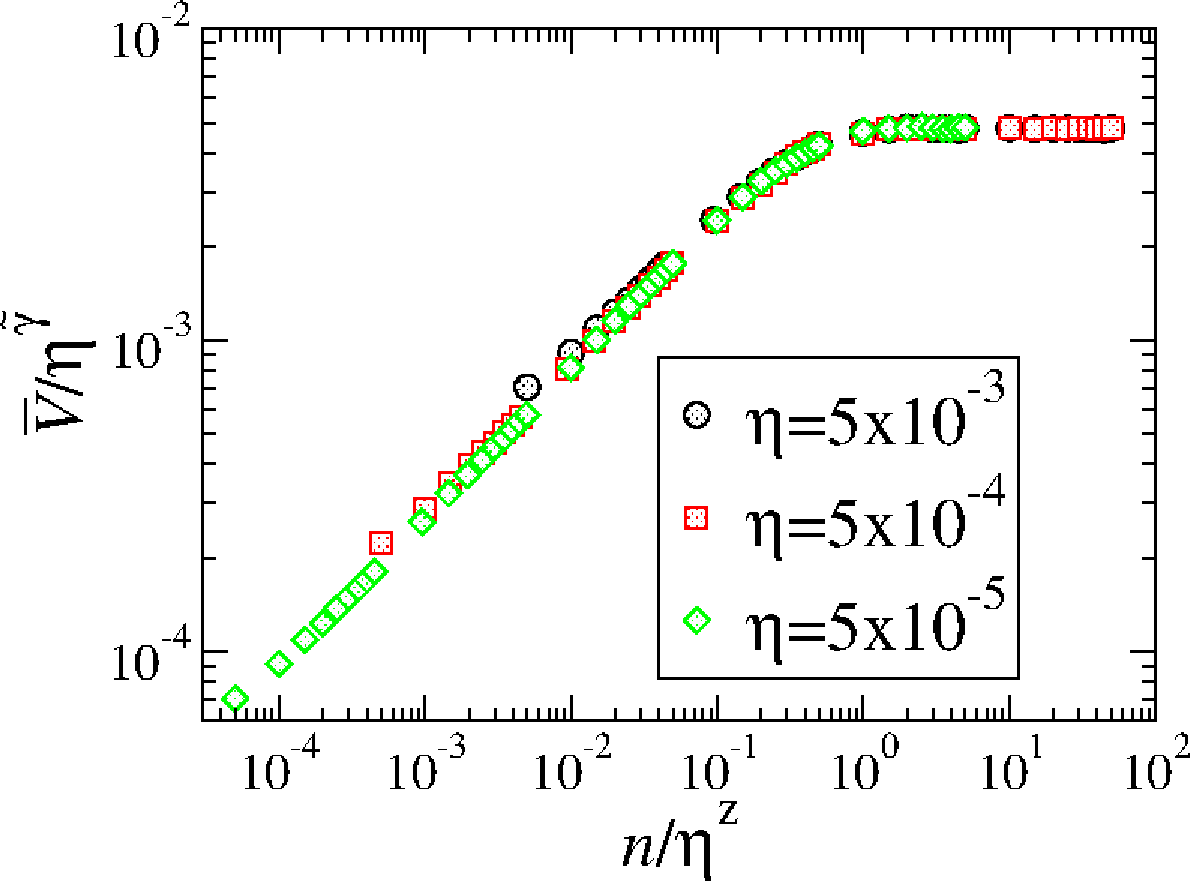}}
\caption{Rescaled axis showing a single and universal plot of three different $\bar{V}$ curves. The control parameters used are shown in the figure.}
\label{aFig6}
\end{figure}

The behavior of the particle's average velocity when a given initial velocity is microscopic compared to the maximum component of the moving boundary velocity is discussed now. A plot of the average velocity as a function of $n$ is shown in Fig. \ref{aFig5}(a).
The average velocity starts growing for small $n$, and after reaching a critical crossover $n_x$, it bends towards a regime of saturation, defined by a constant plateau. As the damping coefficient decreases, the average velocity reaches higher values, and the crossover increases. We describe this behavior considering
\begin{itemize}
\item{For small $n$, say $n\ll n_x$, the average velocity is given by
\begin{equation}
\bar{V}\propto n^{\beta}~,
\label{aeq14}
\end{equation}
where $\beta$ is a critical exponent;}
\item{For very large $n$, i.e. $n\gg n_x$, the average velocity is
written as
\begin{equation}
\bar{V}_{\rm sat}\propto \eta^{\tilde{\gamma}}~,
\label{aeq15}
\end{equation}
and $\tilde{\gamma}$ is also a critical exponent;}
\item{Finally, the crossover $n_x$, which marks the change from the
regime of growth to the saturation is given by
\begin{equation}
n_x\propto \eta^z~,
\label{aeq16}
\end{equation}
where $z$ is a critical exponent.}
\end{itemize}
Using the formalism shown in \cite{re23,re33}, we can utilize a scaling function to describe the behavior of $\bar{V}$. The critical exponents are obtained by numerical fittings as shown in Fig. \ref{aFig5}(b) and Fig. \ref{aFig5}(c), and the obtained values were $\beta=0.4868(5)\cong 0.5$, $\tilde{\gamma}=-0.5005(4)\cong -0.5$ and $z=-1.027(1)\cong -1$. We can rescale the axis using these three values and obtain a single and universal plot, as shown in Fig. \ref{aFig6}.

\subsection{Results for the case $F=-\eta V^{\delta}$}

\begin{figure}[t]
\centerline{\includegraphics[width=0.6\linewidth]{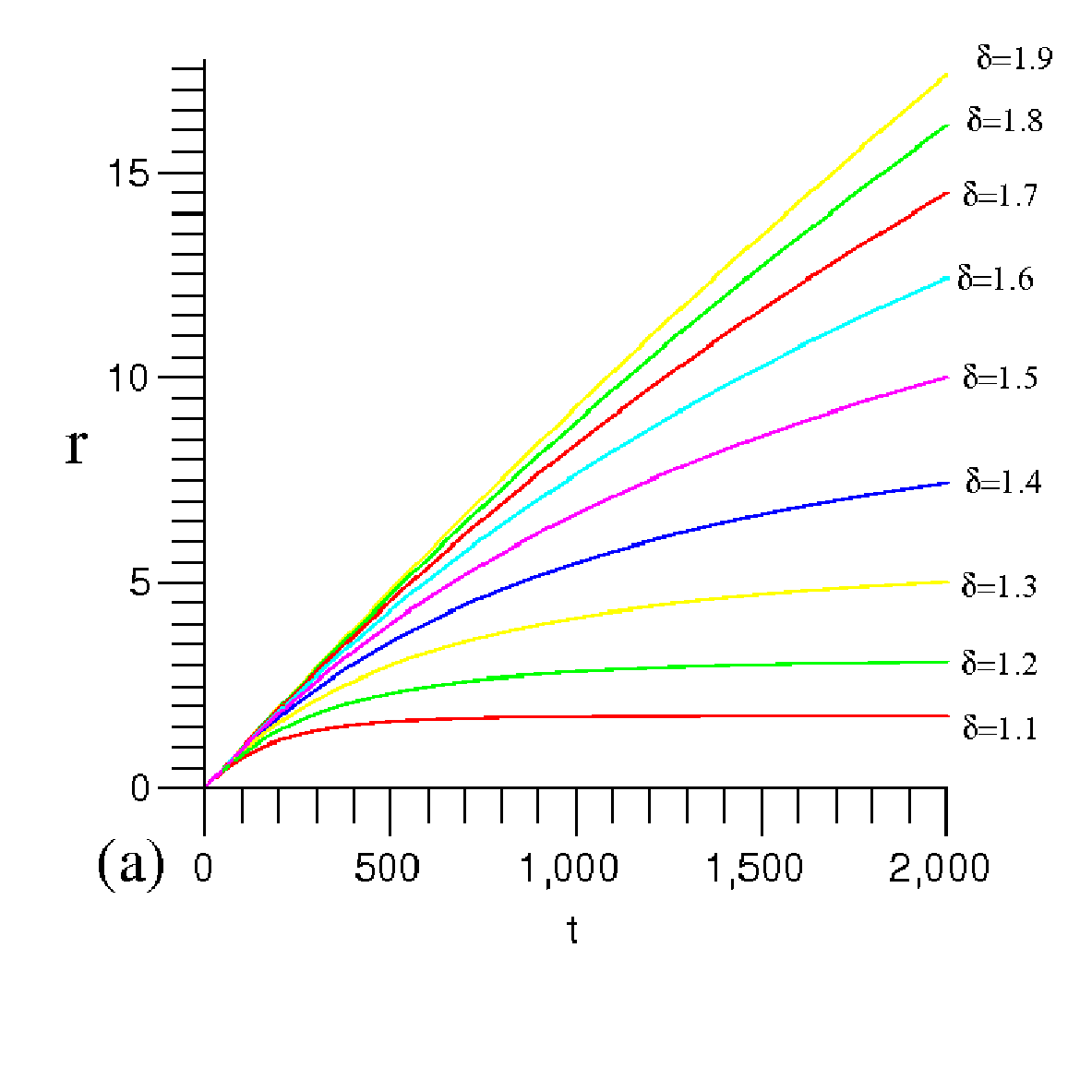}\includegraphics[width=0.6\linewidth]{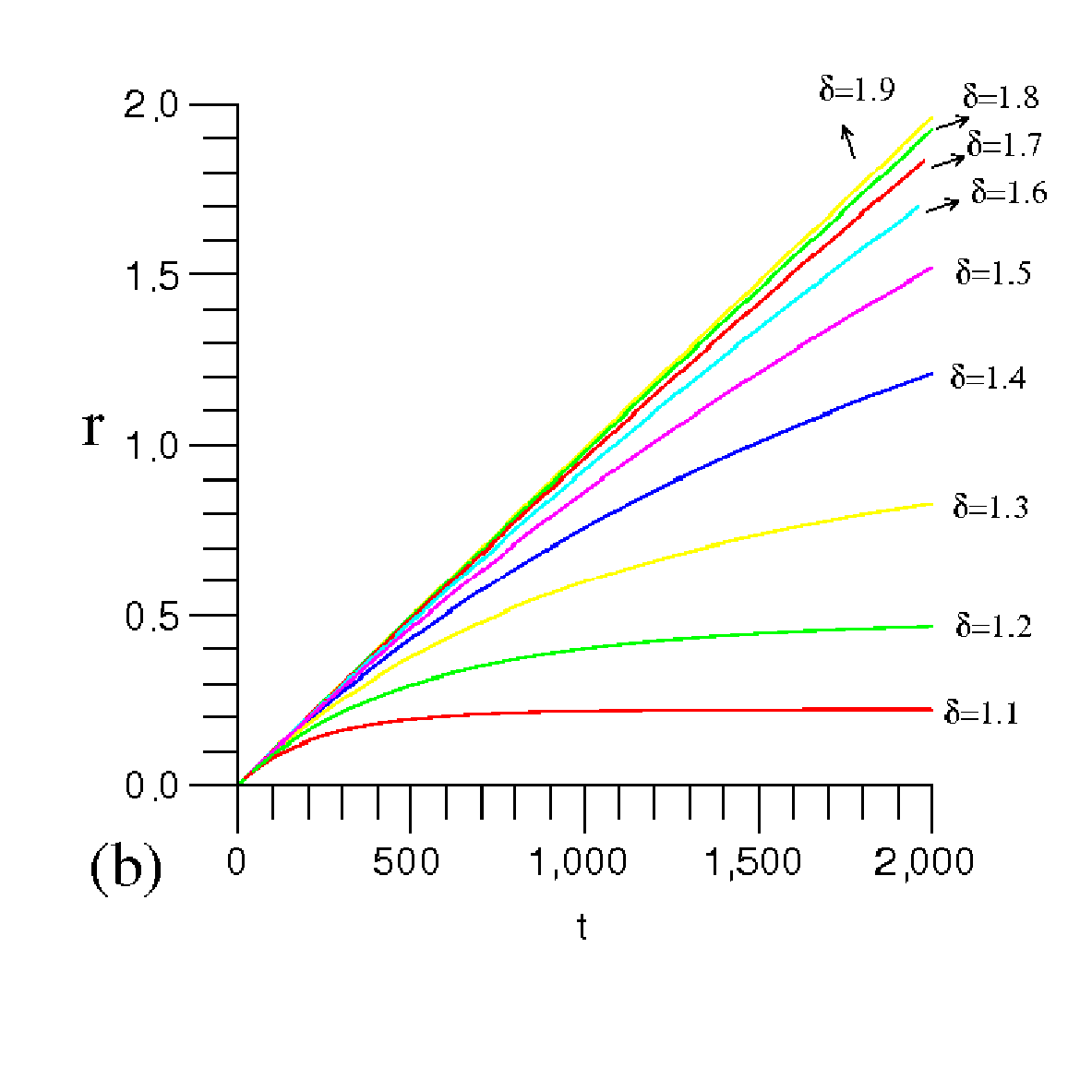}}
\caption{(a) Plot of $r\times t$ for different values of the exponent $\delta$. The initial velocity used was $V_0=10^{-3}$. (b) Same plot of (a) for initial velocity $V_0=10^{-2}$.}
\label{aFig7}
\end{figure}

Let us consider the dissipation $F=-\eta V^{\delta}$ with $1<\delta<2$, acting in the particle. Taking the initial velocity as $V_n>0$ and integrating the equation of motion, we obtain
\begin{equation}
V(t)=[V_n^{\delta}-\eta(1-\delta)(t-t_n)]^{{1}\over{1-\delta}}~,
\label{aeq17}
\end{equation}
with $t\ge t_n$ and $\delta \ne 1$. The displacement of the particle is
obtained by the integration of $dr/dt=V(t)$, yielding
\begin{equation}
r(t)={{V_n^{2-\delta}}\over{\eta(2-\delta)}}-{{\left[
V_n^{1-\delta}-\eta(1-\delta)(t-t_n)\right]^{{2-\delta}\over{1-\delta}}}
\over{\eta(2-\delta)}}~,
\label{aeq18}
\end{equation}
with $\delta \ne 1$, $\delta \ne 2$ and $t\ge t_n$. Depending on the
control parameter $\delta$, the dissipation can lead to a complete
stopping the particle. We illustrate the typical regimes of the
displacement of the particle in Fig. \ref{aFig7}(a) and Fig.
\ref{aFig7}(b)
by some plots of $r\times t$ for $\eta=10^{-2}$ and for two different
values of $V_0$: (a) $V_0=10^{-3}$ and (b) $V_0=10^{-2}$. The
control parameter $\delta$ is shown in the figure. Since the particle may acquire small velocity values in the dynamics, depending on the coordinate angle of the particle's trajectory, all of its energy may dissipate, stopping the dynamics. Such behavior is not observed for $\delta>1.48$ values. See Ref. \cite{re2} for further details.

\section{A discussion on phase transition}

In this section, we delve into the intriguing phase transitions observed in the two systems described throughout the paper originated from a revisitation of Refs. \cite{re1,re2}. Drawing inspiration from the LRA conjecture, which associates chaotic dynamics in a billiard system with fixed boundaries to the emergence of FA under time perturbations.

Both models explored in this study exhibit FA when subjected to a time perturbation at the boundary. The hallmark of FA is the particle's velocity growing as a power law in time, leading to an unbounded diffusion of energy. Our findings reveal a noteworthy phenomenon -- a phase transition from unbounded to bounded growth of energy marked by the introduction of dissipation. This transition mirrors typical phase transitions discussed in statistical mechanics \cite{edl}. We compare the phase transition observed in a ferromagnetic system to provide a practical analogy.

Consider a system of magnetic spins interacting and aligning with an external field. The order parameter, denoted as $m$, signifies the spontaneous magnetization and is a crucial indicator of the system's state. At temperatures below a critical point $T_c$, nonzero magnetization is observed. The ordered phase collapses as the temperature exceeds $T_c$, resulting in zero magnetization. Close to $T_c$, magnetization smoothly approaches zero, exhibiting a divergence in magnetic susceptibility $\chi$.

Identifying a suitable order parameter becomes crucial in our diffusion-driven systems context. FA is closely tied to unlimited energy growth, and our proposed order parameter connects with long-time dynamics, representing a stationary state in the presence of dissipation. For the elliptical billiard under inelastic collisions and similar to what is observed in the time dependent oval billiard (see Ref. \cite{re34}), the saturation of the velocity is described as a power law $V_{sat}\propto (1-\gamma)^{\tilde{\gamma}}$ with $-1<\tilde{\gamma}<0$. The proposed order parameter is $\sigma=1/V_{sat}$ which goes to zero when ${\gamma}\rightarrow 1^{-}$. The susceptibility $\chi=\partial\sigma/\partial\gamma$ diverges as the restitution coefficient $\gamma$ approaches unity.

Turning our attention to dissipation introduced by a drag force, we adopt a similar approach. The saturation of velocity for this case ($V_{sat}$) exhibits power-law behavior ($V_{sat}\propto \eta^{\tilde{\gamma}}$). Defining an order parameter $\sigma=1/V_{sat}$, we find that $\chi=\partial\sigma/\partial\eta$ diverges as the damping coefficient $\eta$ tends to zero.

The observed suppression of FA in both models presents compelling evidence for classifying these phenomena as second-order phase transitions. These findings enrich our understanding of billiard systems' intricate interplay between chaotic dynamics, dissipation, and energy diffusion.

\section{Conclusions}

In summary, mitigating unbounded energy growth in time-dependent billiards, attributed to dissipative forces, is a complex phenomenon influenced by different mechanisms. The impact of inelastic collisions and the imposition of a drag force both play roles in curbing the uncontrolled energy diffusion within an ensemble of particles navigating a time-dependent billiard system. Establishing a steady-state velocity over extended periods provides compelling evidence for this restraining effect. Notably, the diverse nature of dissipative forces highlights the non-robust nature of Fermi acceleration in billiards, as demonstrated by the observed velocity saturation persisting over prolonged intervals.

\section*{Acknowledgements}

EDL thanks CNPq (National Council for Scientific and Technological Development, Brazil) for the financial support from grants 301318/2019-0 and 303707/2015-1. Additional support from FUNDUNESP and FAPESP (S\~ao Paulo Research Foundation) through grants 2021/09519-5, 2019/14038-6, 2017/14414-2, 2012/23688-5, 2008/57528-9, and 2005/56253-8 is also acknowledged.

We also acknowledge the support from the Center for Scientific Computing (NCC/GridUNESP) at S\~ao Paulo State University (UNESP), which provided essential research resources.

\end{document}